\documentclass[a4paper]{amsart}
\usepackage{amsmath}
\usepackage{amssymb}
\usepackage{bbm}
\usepackage{a4wide}
\usepackage{graphicx}
\usepackage{color}

\parskip\smallskipamount

\let\set\mathbbm
\def\<#1>{\langle#1\rangle}

\def\e{\mathrm{e}}
\def\e1{\varepsilon}

\begin{document}

\author[Christoph Koutschan]{Christoph Koutschan\,$^2$}
\address{Christoph Koutschan, Research Institute for Symbolic Computation, Johannes Kepler University Linz, Austria, 
         and Tulane University, New Orleans, LA, USA (from October 2009 until June 2010).}
\email{ckoutsch@risc.jku.at}

\author[Doron Zeilberger]{Doron Zeilberger\,$^3$}
\address{Doron Zeilberger, Mathematics Department, Rutgers University (New Brunswick), Piscataway, NJ, USA.}
\email{zeilberg@math.rutgers.edu}

\thanks{$^1$The final publication is available at www.springerlink.com, DOI: 10.1007/s00283-010-9192-1.}
\thanks{$^2$supported in part by NFS-DMS 0070567, and by the Austrian Science Fund (FWF): P20162-N18}
\thanks{$^3$supported in part by the United States of America National Science Foundation}

\title[The Pekeris-Accad-WEIZAC Ground-Breaking Collaboration]
{The 1958 Pekeris-Accad-WEIZAC Ground-Breaking Collaboration 
that Computed Ground States of Two-Electron Atoms
(and its 2010 Redux)$^1$}

\maketitle
\setcounter{footnote}{3}
\thispagestyle{empty}

\section*{We have come a long way}

In order to appreciate how well off we mathematicians and scientists are today,
with extremely fast hardware and lots and lots of memory, as well as with
powerful software, both for numeric and symbolic computation, 
it may be a good idea to go back to the early days of electronic computers and compare how things went then.
We have chosen, as a case study,
a problem that was considered a huge challenge at the time.
Namely, we looked at C.L. Pekeris's~\cite{Pek} seminal 1958 
work\footnote{available on-line from {\tt http://astrophysics.fic.uni.lodz.pl/100yrs/pdf/04/076.pdf} 
(viewed May 15, 2010)} on the ground state energies of two-electron atoms.
We went through all the computations {\it ab initio} with today's software and hardware.

\section*{Schr\"odinger}

Let's recall the (time-independent) Schr\"odinger equation for the {\it state function} (alias {\it wave function})
$\psi(x,y,z)$ of a one-electron atom with a stationary nucleus  
(see, for example, \cite{PW} Eq.~(30-1) with $N=1$), in {\it atomic units}:
\[
  \left ( \frac{\partial^2 }{\partial x^2}+\frac{\partial^2 }{\partial y^2}+\frac{\partial^2 }{\partial z^2}+
  2 \left ( E+ \frac{Z}{r} \right )  \right ) \psi (x,y,z) \; = \; 0,
\]
where~$Z$ denotes the nuclear charge, $E$ the energy of the
system, and $r=\sqrt{x^2+y^2+z^2}$ the distance
of the electron to the nucleus.

Schr\"odinger's solution of this eigenvalue problem is one of the greatest classics of modern physics,
familiar to all physics students (and chemistry students, but unfortunately not math),
using separation of (dependent) variables, and getting explicit and exact results
for the eigenvalues (the possible energy levels~$E$) and even for the 
corresponding eigenfunctions~$\psi$.
Because the eigenfunctions (or more precisely their squares) are interpreted as probability
distributions, certain restrictions have to be imposed on~$\psi$; in particular,
the integral of $|\psi|^2$ over the whole domain must be finite. The eigenvalues
then are exactly those values of~$E$ for which the Schr\"odinger equation
admits such a solution.
It turns out that these eigenfunctions are expressible in terms of the venerable special functions of
mathematical physics, namely (associated) Legendre and (associated) Laguerre polynomials.

But exactly the same predictions (about the energy levels) were already made by the ``old'', {\it ad hoc}, Bohr-Sommerfeld quantum mechanics;
the ``new'' wave- and matrix-quantum theories needed to predict facts that were beyond the scope of the old theory,
thereby offering a crucial confirmation.
That's why Schr\"odinger himself, Hylleraas, and many other physicists tried to derive the
energy levels (alias eigenvalues) for two-electron atoms, whose Schr\"odinger equation, for the
wave function $\psi=\psi(x_1,y_1,z_1,x_2,y_2,z_2)$, is
\[
  \left( \frac{\partial^2 }{\partial x_1^2}+\frac{\partial^2 }{\partial y_1^2}+\frac{\partial^2 }{\partial z_1^2}+
  \frac{\partial^2 }{\partial x_2^2}+\frac{\partial^2 }{\partial y_2^2}+\frac{\partial^2 }{\partial z_2^2}+
  2 \left ( E+ \frac{Z }{r_1}+ \frac{Z}{r_2} -\frac{1}{r_{12}} \right ) \right ) \psi \; = \; 0,
\]
where~$E$ and~$Z$ are as above, while $r_1$, $r_2$ are the distances of the electrons from the nucleus,
and $r_{12}$ is their mutual distance.

The task turned out to be forbidding.
There were some crude attempts to use perturbation theory, but none of their predictions came
close to the experimental spectra already known then. It was a major challenge to
vindicate the new quantum mechanics by computation. For once, 
the experimenters were ahead, and the theorists had
to catch up.

\section*{Pekeris}

Chaim Leib Pekeris (1908--1993) had a brilliant idea how to catch up. With a computer, of course!
He had a carefully laid-out approach, to be described soon, that would indeed give a very
accurate prediction of the helium spectrum, given a powerful enough computer and
a clever enough programmer.

Except that when he first had that idea, computers didn't yet exist, and when finally 
he had access to the JOHNNIAC, during his frequent long visits to the
Institute for Advanced Study up to von Neumann's death (in 1957),
it was not quite powerful enough, and at any rate was too busy, to pursue Pekeris's plan.

In addition to being a brilliant scientist, Pekeris was also an ardent Zionist.
His good friend (another Chaim, and another scientist), Chaim Weizmann (1874--1952),
invited him, already in 1947, to head the department of applied mathematics
at the Ziv Institute (later renamed the Weizmann Institute of Science),
and Pekeris agreed---  in principle, but only on condition that they build
a computer similar to the JOHNNIAC. A committee was formed, including no lesser 
figures than Albert Einstein and John von Neumann, to decide whether this
was a good idea. Einstein believed not. In those days computers were very
expensive, and he thought that such a poor, developing country could make better
use of such a big chunk of money; but von Neumann managed to win Einstein over and
the plan was approved. It took a few years to materialize, and finally they recruited
one of the members of von Neumann's team, a visionary electrical engineer by the name
of Gerald Estrin (b. 1921)~\cite{Est}. Estrin recounts (\cite{Est}, p. 319) that
in one short conversation with von Neumann, shortly before his departure,
he asked, ``What will that tiny country do with an electronic computer?'' John
von Neumann responded: ``Don't worry about that problem. If nobody else uses the computer,
Pekeris will use it full time!'' Estrin comments that this turned out to be
an important prophecy that he often recalled.
\begin{center}
 \parbox[t]{40mm}{
  \begin{center}
  \includegraphics[height=50mm]{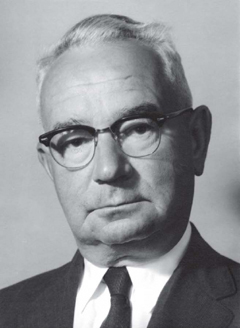}\\ C. L. Pekeris$^5$ 
  \end{center}
 }
 \parbox[t]{70mm}{
  \begin{center}
  \includegraphics[height=50mm]{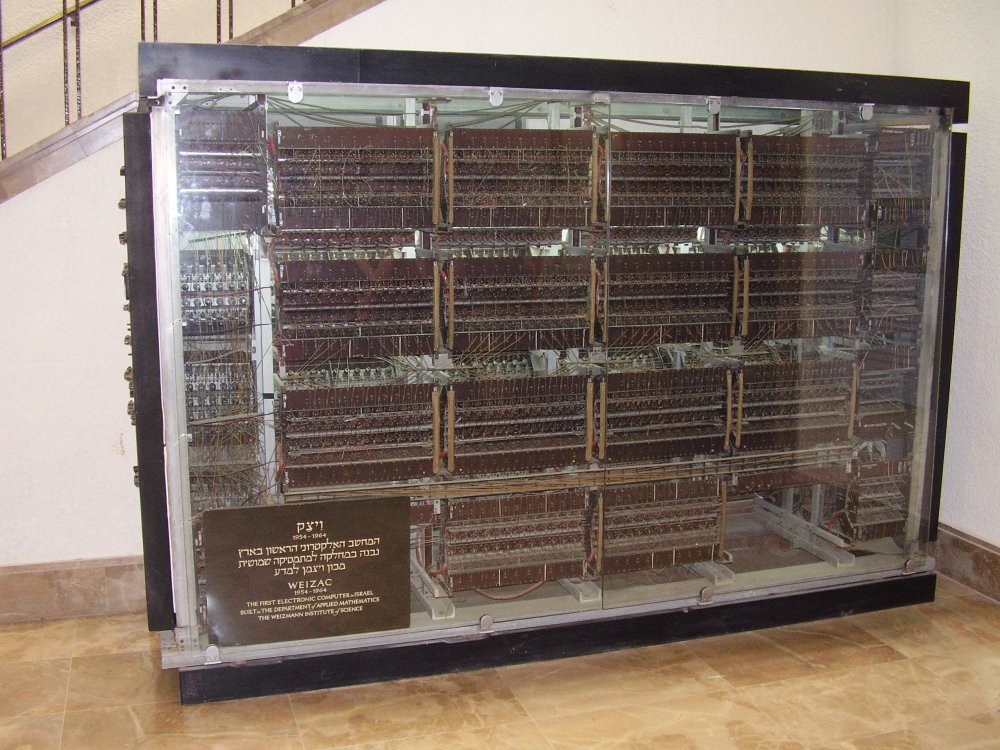}\\ WEIZAC$^6$ 
  \end{center}
 }
\end{center}
\textcolor{white}{  
  \footnote{Photo courtesy of Optik Foto Rutz AG, St. Moritz, Switzerland}
  \footnote{Photo courtesy of Yuval Madar under the Creative Commons Attribution ShareAlike 3.0 License,
            taken from {\tt http://en.wikipedia.org/wiki/WEIZAC}}
}

\section*{Pekeris's Crazy Plan}
 
The first step was standard.  Using the symmetries of the problem, one sees  
that the wave function~$\psi$ of the {\it ground state} depends only on
$r_1,r_2,r_{12}$, so one ``merely'' has to deal with functions of three variables,
rather than six.
The new partial differential equation,
in variables $r_1,r_2,r_{12}$, is easily derived (Eq.~(5) of~\cite{Pek}).

The next step (first suggested by H.M. James and A.S. Coolidge, see ref. 4 of~\cite{Pek})
was to make another change of variables, this time a linear one. After substituting $E=-\e1^2$,
introduce  {\it perimetric coordinates}:
\begin{eqnarray*}
  u & = & \e1(r_2+r_{12}-r_1),\\
  v & = & \e1(r_1+r_{12}-r_2),\\
  w & = & 2\e1(r_1+r_2-r_{12}).
\end{eqnarray*}
These new variables have the advantage that they range freely and independently from $0$ to $\infty$.
In contrast, $r_1$, $r_2$, and~$r_{12}$ are the lengths of the sides of a triangle 
(whose vertices are the two electrons and the nucleus),
and so must obey the triangle inequality. 
In addition, the expected asymptotic behavior of~$\psi$, deduced from
the hydrogen (one-electron) case, suggested writing (\cite{Pek}, Eq.~(13))
\[
  \psi=e^{-\frac{1}{2} (u+v+w)} F(u,v,w),
\]
and letting $F(u,v,w)$ be the function sought. Pekeris performed this change of variables
---purely by hand--- and derived a fairly hairy linear partial differential equation
with polynomial coefficients, satisfied by~$F$, that we do {\it not} reproduce here; the curious reader
can either look it up (\cite{Pek}, Eq.~(14)), or look at the computer output that
is available from our webpages. 

The next step was to express $F(u,v,w)$ as a series expansion of products
of (simple) Laguerre polynomials (Eq.~(16) of~\cite{Pek}):
\[
  F \, = \, \sum_{l,m,n=0}^{\infty} A(l,m,n) L_l(u) L_m(v) L_n(w),
\]
where $L_n(x)$ denotes the Laguerre polynomial
\[
  L_n(x) \, = \, \sum_{k=0}^{n} \binom{n}{k} \frac{(-x)^k}{k!}.
\]
Like all families of classical orthogonal polynomials, the Laguerre polynomials satisfy
a pure (linear) differential equation, a pure (linear) recurrence equation,
and a mixed differential-recurrence relation:
\begin{eqnarray*}
  xL_n''(x) & = & (x-1)L_n'(x)-nL_n(x),\\
  xL_n(x) & = & -(n+1)L_{n+1}(x)+(2n+1)L_n(x)-nL_{n-1}(x),\\
  xL_n'(x) & = & nL_n(x)- nL_{n-1}(x),
\end{eqnarray*}
the primes denoting differentiation with respect to~$x$.

Now came an astounding feat! Pekeris substituted the expansion for~$F(u,v,w)$, in terms
of the yet-to-be-determined $A(l,m,n)$, into the above-mentioned linear differential
equation (Eq.~(14) of~\cite{Pek}, politely not shown here), and using the
above relations for the Laguerre polynomials got rid of all differentiations, and
then, by using the pure recurrence, got rid of any monomials in $u,v,w$. Then he collected terms,
and got ---purely by hand--- a huge {\it monster}, a 33-term linear partial recurrence
equation with polynomial coefficients satisfied by the $A(l,m,n)$.
Each of the coefficients of the $33$ shifts $A(l+\alpha,m+\beta,n+\gamma)$ that showed up was
polynomial in $l,m,n$ of degree $3$, and of degree $1$ in in the charge~$Z$ and the yet-to-be found~$\e1$.

We will kindly spare the reader this recurrence (and spare ourselves
from typing it!), but the truly courageous reader can glance at Eq.~(22) of~\cite{Pek}.
 We shudder to think of the pain of the poor typist who keyed this from Pekeris's
hand-written manuscript, and the type-setter for {\it Physical Review}, 
not to mention Pekeris himself.
They all deserve lots of credit.
In his  wonderful essay~\cite{Est} (p. 331), Estrin understates the point:
the ``appearance of this ugly $33$-term recurrence would be enough to discourage most analysts.''

The recurrence yielded a homogeneous linear system of equations with $\infty^3$ equations and $\infty^3$ unknowns,
that usually has no non-trivial solutions, but for some~$\e1$, the ``eigenvalues'', the ``determinant vanishes''
and there are solutions. The largest eigenvalue is of primary physical
relevance, for it corresponds to the ground state energy of the atom.

But even the most powerful computers can handle only finite systems! Hence the next step 
consisted in reducing to a finite, truncated version of the system, considering only those
$l,m,n\geq0$ for which $l+m+n \leq \omega$, for some finite~$\omega$ and setting all the
$A(l,m,n)$ with $l+m+n > \omega$ equal to~$0$. In addition, the system could be cut approximately in half
by requiring either symmetry ($A(l,m,n)=A(m,l,n)$, the so-called {\it para states})
or antisymmetry ($A(l,m,n)=-A(m,l,n)$, the so-called {\it ortho states}).

If this was to be handled on a computer (even one which did not yet exist), one needed
a convenient way to order linearly all the triplets of integers $(l,m,n)$ with $l+m+n \leq \omega$ and
$l \leq m$ in the symmetric case (resp. $l<m$ in the antisymmetric case). 
For this Pekeris devised a fairly complicated bijective map 
$k\colon\{(l,m,n)\in\set{N}_0^3\mid l\leq m\}\to\set{N}$ which once
again we spare the reader, but which can be found in Eqs.~(27-29) of~\cite{Pek}
(by the way, Eq.~(28) contains a very rare misprint, there should be $\frac{1}{2}(l+m)$ added to it).

It is not known when Pekeris devised this plan, but it was probably several years before
he had access to a computer, so he just had to wait until Chaim Weizmann's promised computer
would materialize, carrying out the recommendation of the above-mentioned committee of Einstein, von Neumann et al.
The difficulty of the problem that Pekeris faced becomes even more
evident when taking into account that some closely related problems
are still open.  For example, it is experimentally known that all
existing atoms can form negative ions with no more than one or two
extra electrons, but there is no theoretical understanding of this
phenomenon.

\section*{WEIZAC}

We have already mentioned Estrin,
the person chosen to head the team that would build from scratch the first Israeli electronic
computer, and highly recommended his vivid account~\cite{Est}. The WEIZAC team consisted of a cadre of young and talented
electrical engineers (including Aviezri Fraenkel (b. 1929) who later 
did a Ph.D. in number theory, 
became, inter alia, an authority on combinatorial games, and pioneered the use of computers in religious studies).

Finally the computer was ready, and Pekeris was itching to use it on his many problems, including
the spectrum of helium, but he needed a {\it programmer} (what today we would call a ``software engineer'', but
there was no such thing as software in those days).  Not,
of course, a Java programmer, nor a Fortran programmer,
and not even an Assembly-language programmer. Back in 1957 these were yet to be invented. The
only language that WEIZAC understood then was {\it machine language}, and the alphabet consisted of
two letters only, $0$ and $1$ (via the 16-letter alphabet of hexadecimals).
But how to find such programmers? Definitely not among
graduates of computer science departments, for there were none.

What Pekeris did was ask his secretary to place classified ads in the daily newspapers, asking for high school graduates,
after their military service, who attended the {\it megama re'alit} (math/science track).

\section*{Accad}

Yigal Accad (b. 1936), fresh out of his military service,  answered such an ad. 
In a recent e-mail message, dated May 7, 2010, Accad recalls:
\begin{quote}
On a 1957 Friday (or was it a Holiday Eve) that happened to be a
non-working day at the Weizmann Institute, Prof. Pekeris unexpectedly drove
his 1948 Studebaker to our residence at the southern edge of Rehovot. He
invited me to join him in his office. Over there he pulled out a pile of
handwritten papers and went with me through many of the equations you can
find in the 1958 paper, including Eq.~(22). As I remember, this tour took at
least 2 hours. At the end Prof. Pekeris asked me if I can handle this
problem. There were only 2 possible answers to this question and the rest is
history. This may have been the best risk I have taken.
\end{quote}

Estrin goes on to state the following  accolades (\cite{Est}, p. 330):
\begin{quote}
There is a clear testimony to the fact that Yigal Accad had unusual ability to use
WEIZAC as a tool with very little software between him and the machine semantics.
That ability, when combined with his talents as an applied mathematician, was a significant
factor in the ensuing problem-solving successes at the Weizmann Institute.
\end{quote}

Accad became Pekeris's right-hand man for many years, and it is hard to imagine what Pekeris
would have done without him. Pekeris appreciated Accad's invaluable work,
and it was at his suggestion that Yigal, while
working full-time as a software engineer, enrolled in the graduate school
(after completing his undergraduate studies at Hebrew University)
and incorporated some of the research into, first a master's thesis, in 1969, and then 
in 1973 a Ph.D. thesis, which was a far-reaching extension of the work we recount here). 

Accad stayed at the Weizmann Institute from 1956 until 1989. Between 1977 and 1989 he also
served as a consultant to the pioneering Israeli Hi-Tech company Scitex. 
In 1989 he moved to California and joined Electronics for Imaging (EFI),
working there until 2008, ultimately becoming chief scientist.

\section*{The Pekeris-Accad-WEIZAC collaboration}

Indeed Accad was the perfect person to tame Pekeris's monster recurrence, to 
write (machine-language) programs to generate
the truncated matrices, and to implement the iterative algorithm for estimating the largest
eigenvalue.
The impressive (for its time) WEIZAC output is displayed in
Table III of~\cite{Pek} for values of the charge~$Z$ ranging from $Z=1$ to $Z=10$.
We are happy to report that our 2010 computations (on three different platforms)
completely agree with that table, all the way to the last decimal digit.

In a follow-up paper, published a year later, Pekeris~\cite{Pek1} 
(and of course, Accad and WEIZAC---but it would be more than 30 years later
before any computer, Shalosh B. Ekhad, became co-author of a published paper!)) 
treat the important special case of helium ($Z=2$) with a greater accuracy,
and also consider the ortho state $2 \,\, {}^{3}S$. 
Our computations agree with that paper, too.

\section*{2010}
Of course, thanks to Moore's Law, all these computations can now be done
much faster, and there is no reason for us to be proud
that we can compute the eigenvalues within seconds with today's
hardware and software, a task that kept WEIZAC busy round-the-clock for months:
for example, a fixed-point multiplication took 1 millisecond on this
early computer and the capacity of its memory was 4096 words (40 bits per word).
But what is still remarkable and probably not so obvious:
not only the WEIZAC part, the numeric computation that can now be done
on every laptop, and the Accad part, challenging in machine language but
today an easy exercise with high-level programming
languages, but also, and {\it especially}, the Pekeris part
can now be done much faster and mostly automatically, using computer algebra. 
Even more: in view of the gigabyte-sized recurrences that we can currently handle
(see for example~\cite{KoutschanKauersZeilberger10})
with symbolic software, the ``monster recurrence'' looks rather dwarfish.
We don't know exactly how long it took Pekeris to derive the differential equation
and the recurrence, but let's say 20 person-hours (including
checking and rechecking); our program needs 0.108 seconds. 

To be honest, it took us a couple of hours to {\it program} Maple and
Mathematica to follow Pekeris's plan, but with almost the same effort,
one could (and we did) program the {\it general problem}, that could
be used again and again for many other differential equations in
future problems.  Our programs {\tt PEKERIS} (for Maple, by DZ) and
{\tt Pekeris.nb} (for Mathematica, by CK) are indeed very general:
they basically can input {\it any} linear differential equation, in
{\it any} number of variables, and {\it any} series of substitutions,
and output the transformed differential operator. Also the recurrence
for a Laguerre polynomial expansion is achieved completely
automatically.  Using the widely known concept of Gr\"obner bases 
(invented by Bruno Buchberger in 1965 and hence
not yet available for Pekeris) it is also possible to perform the
series expansion for any set of orthogonal polynomials of
hypergeometric type. For this purpose, the defining equations for the
family of polynomials are represented as a Gr\"obner basis, which
makes sense when the relations are rewritten, in operator notation, as
(noncommutative) polynomials. Having chosen an appropriate monomial
order, the elimination of the differentials can be achieved by a
simple reduction modulo the Gr\"obner basis. Similarly, by changing
the underlying polynomial ring, the elimination of the continuous
variables $u,v,w$ can be done. Let us also remark that you don't need to be a
Laguerre or a Pekeris to generate the relations for the Laguerre (and
other orthogonal) polynomials.  They are all routinely derivable (and
provable) by the so-called Wilf-Zeilberger method~\cite{PWZ}, as 
implemented, e.g., in the Mathematica package {\tt HolonomicFunctions}~\cite{Koutschan09}
that we employ in our program.

Modular techniques using Chinese remaindering and polynomial
interpolation allow for computing the determinant {\it symbolically}
up to quite large dimensions: for example, the determinant of the
$161\times161$ matrix (para case with $\omega=10$) is obtained in less
than five minutes, yielding a polynomial in~$\e1$ of degree~$161$
having integer coefficients with about $500$ digits! It is clear that
this strategy produces a lot of overhead, so that an alternative way
is desirable. We reformulate the problem of finding the largest~$\e1$ 
for which the determinant of $M\in\set{Z}[\e1]^{n\times n}$ vanishes,
as a {\it generalized eigenvalue problem}:
\[
  Av = \e1 Bv,\quad M=A-\e1 B\text{ with }A,B\in\set{Z}^{n\times n}.
\]

Although Maple and Mathematica are computer algebra systems for symbolic computations in the first
place, they also offer quite some functionality for numerical computations,
in particular for the above problem. But since we were not 100\% satisfied
with either---Maple was rather slow for the desired precision and Mathematica
didn't allow higher precision than machine reals (6 decimal digits)---we
tried with MATLAB, a software designated for numeric computations,
especially in linear algebra.
Notably, the program code for building the (sparse) matrices is itself
computer-generated! It contains the $33$ terms of the recurrence 
{\it hard-coded} to produce the matrix entries, and therefore certainly
comes closer to Accad's machine-code program. We were very impressed
by MATLAB's speed and accuracy. Computing all entries of Table III of~\cite{Pek}
takes less than a second, and without much effort $\omega$ can be increased
to $60$, corresponding to a $20336\times20336$ matrix.

\section*{Software and Sample Output}

This article is accompanied by the Maple package {\tt PEKERIS}, available from
\begin{center}
\verb|http://www.math.rutgers.edu/~zeilberg/mamarim/mamarimhtml/pekeris.html| \quad ,
\end{center}
where the reader can also find lots of output files (and input files if they want to modify them
to get more output) that reproduce and far extend the seminal 1958 computations of
Pekeris, Accad, and WEIZAC.
Further we provide the Mathematica notebook {\tt Pekeris.nb} (for which the package {\tt HolonomicFunctions}
is required), and the MATLAB programs {\tt PekerisPara.m} and {\tt PekerisOrtho.m}, 
all available from
\begin{center}
\verb|http://www.risc.jku.at/people/ckoutsch/pekeris/| \quad .
\end{center}
Our maplephone readers are welcome to play with the first package while
the mathematicaphones would probably prefer the latter one.
However, even people (shame on you!) who speak neither Maple nor Mathematica can appreciate
the output files, written in plain humanese. The second-named author is particularly proud of the procedure
{\tt PaperPara} that fully automatically and seamlessly generates a whole
article, ready to be submitted to {\it Physical Review}, without any human touch.
Changing the parameters can produce many similar papers, see
\begin{center}
\verb|http://www.math.rutgers.edu/~zeilberg/tokhniot/oPEKERIS1| \quad .
\end{center}

\section*{Conclusion}

This article is first and foremost an ode to the vision and ingenuity of computing pioneers,
but it also makes the point that there are lots of hidden treasures in the ``old'' 
scientific literature, that can be revisited with today's powerful symbolic computation software.
We are not the first to advocate using symbolic computations in scientific computing,
see for example~\cite{Bar} (unfortunately he was unaware of~\cite{WZ}), 
and the current impressive application to high-energy physics~\cite{BBKS},
but we believe that there is a huge potential for exploiting symbolic computation on problems
that previously seemed intractable.
This would complement the extensive use
(and according to Nobelist Philip Anderson, excessive and sometimes abusive use ~\cite{And})
of Monte Carlo methods. In particular, the Wilf-Zeilberger algorithmic proof theory~\cite{PWZ}
(and more importantly the subsequent generalizations to multi-summation and multi-integration
\cite{WZ,AZ}), should be taught to all scientists. We would be more than happy if this
article could seed future collaborations between symbolic computation and physics,
chemistry, or other sciences.

\section*{Encore}

Many people, even today, are not comfortable with computer-generated or even computer-assisted proofs,
like the four-color theorem or the Kepler conjecture: they are uncomfortable trusting
the computer. While the ``monster recurrence'' discussed above was still derived purely by
hand, Pekeris must have started using his own ``symbolic'' computation when he tackled 
seemingly intractable problems.
Let us end with his prophetic words (\cite{Pek1960}, quoted in~\cite{Est}, p. 333):
\begin{quote}
Here we are confronted with problems where the computer writes the formulae as well
as evaluates them. By the nature of their origin such formulae are very long---in many cases
too long to be published. We shall therefore be dealing in the future with equations
which only the computer will see. The prospect of operating with invisible equations
is a frightening one, but the alternative is to accept the situation of the past, where problems
have been staring at the applied mathematician for decades, and even more for centuries, without
a practical solution being reached. A problem, like the tides of the oceans, for example,
is not necessarily insoluble just because it had remained in the books for 184 years.
\end{quote}

\end{document}